\documentclass[fleqn,10pt]{wlscirep}
\usepackage[utf8]{inputenc}
\usepackage[T1]{fontenc}
\usepackage{amsmath}
\newtheorem{theorem}{Theorem}
\newtheorem{proposition}{Proposition}
\usepackage{amssymb}
\usepackage{physics}
\usepackage{graphics}
\usepackage{booktabs,multirow}
\usepackage{caption}
\usepackage{hyperref}
\usepackage{color}
\usepackage{booktabs}
\usepackage{multirow}
\usepackage{subcaption}
\usepackage{algorithm}
\usepackage{algpseudocode}
\usepackage{etoolbox}
\makeatletter
\patchcmd{\algorithmic}{\small}{}{}{} 
\makeatother

\usepackage{caption}
\title{A Neural-Guided Variational Quantum Algorithm for Efficient Sign Structure Learning in Hybrid Architectures}

\usepackage{hyperref}
\usepackage{qcircuit}
\author[1,2$\dagger$]{Mengzhen REN}
\author[3$\dagger$]{Yu-Cheng Chen}
\author[4,5*]{Yangsen Ye}
\author[3*]{Min-Hsiu Hsieh}
\author[2,6*]{Alice Hu}
\author[1*]{Chang-Yu Hsieh}
\affil[1]{Zhejiang University, College of Pharmaceutical Sciences, Hangzhou, China}
\affil[2]{Department of Mechanical Engineering, City University of Hong Kong, Hong Kong SAR, China}
\affil[3]{Hon Hai Research Institute, Taipei, Taiwan}
\affil[4]{Hefei National Research Center for Physical Sciences at the 
Microscale and School of Physical Sciences, University of Science and 
Technology of China, Hefei 230026, China}
\affil[5]{Shanghai Research Center for Quantum Science and CAS Center for 
Excellence in Quantum Information and Quantum Physics, University of Science 
and Technology of China, Shanghai 201315, China}
\affil[6]{Department of Materials Science and Engineering, City University of Hong Kong, Kowloon, Hong Kong SAR, China}

\affil[1*]{kimhsieh@zju.edu.cn}
\affil[2,6*]{alicehu@cityu.edu.hk}
\affil[3*]{min-hsiu.hsieh@Foxconn.com}
\affil[4,5*]{ysye@ustc.edu.cn}
\affil[$\dagger$]{these authors contributed equally to this work}

\begin{abstract}
Variational quantum algorithms are a leading candidate for near-term quantum computing, yet their scalability is hindered by high measurement costs, barren plateaus, and noise sensitivity. These challenges stem largely from requiring quantum circuits to learn both probability amplitudes and phase (sign) structure simultaneously—a demanding regime that increases circuit depth and sampling overhead. Here we introduce sVQNHE, a hybrid quantum–classical framework that explicitly separates these roles: a classical neural network models the amplitude distribution, while shallow, commuting diagonal quantum circuits learn only the phase structure. Furthermore, sVQNHE employs an iterative amplitude-transfer mechanism that progressively integrates the classical distribution into the quantum ansatz—while leveraging commuting diagonal gates to reduce gradient measurement cost to $\mathcal{O}(1)$ per operator—ensuring the neural network models only a residual correction on a highly trainable energy landscape with polynomially bounded gradient variance. Across frustrated quantum spin models, molecular Hamiltonians, and combinatorial optimization benchmarks, sVQNHE significantly outperforms conventional variational quantum eigensolvers with comparable parameter counts, converging faster and demonstrating improved robustness under finite sampling and noise. Notably, using qubit-efficient encodings, sVQNHE matches the state-of-the-art classical Free-Energy Machine on a 1,485-vertex MaxCut instance using Pauli Correlation Encoding, and substantially outperforms all baselines on Maximum Clique problems with 135 vertices. These results establish that explicit amplitude–phase separation, with quantum resources dedicated mainly to learning phase structure, provides a principled and scalable design strategy for hybrid quantum–classical algorithms under near-term noisy hardware constraints.
\end{abstract}

\begin{document}

\maketitle
\section*{Introduction}
Hybrid quantum–classical approaches represent a promising strategy for extracting computational value from noisy intermediate-scale quantum (NISQ) and the subsequent early fault-tolerant quantum computing (Early FTQC) devices\cite{preskill2018quantum,katabarwa2024early,nanda2024explorative}. Central to this paradigm is the idea of variational wavefunction learning, in which a parametrized ansatz is iteratively optimized to minimize a task-specific objective, such as ground-state energy or a combinatorial cost function\cite{peruzzo2014variational,mcardle2020quantum,cerezo2021variational}. Despite substantial progress across physics, optimization, and quantum machine learning, current variational quantum algorithms (VQAs) scale poorly in the presence of noise and sampling constraints\cite{cerezo2021variational,fedorov2022vqe}. In practice, training often becomes unstable as system size grows, the gradient signal rapidly diminishes, and measurement cost increases with circuit complexity\cite{cerezo2021cost, grant2019initialization, holmes2021barren, holmes2022connecting, wang2021noise,ilin2024dissipative}.

These failures stem from a profoundly imbalanced division of labor between quantum and classical resources. In most contemporary VQA architectures, from hardware-efficient layers to QAOA-style circuits, the quantum processor is often tasked to optimize a generic ansatz state with both the probability amplitudes and the delicate phase/sign structure to be tuned simultaneously. Capturing realistic amplitude distributions typically demands highly entangling gates and considerable circuit depth\cite{grimsley2019adaptive,chen2020variational}, whereas correctly learning the phase is extraordinarily fragile under hardware noise\cite{cerezo2021cost,wang2021noise,ragone2024lie}. As a result, improvements in one aspect often degrade the other, driving the well-documented trade-offs among circuit depth, trainability, sample complexity, and noise sensitivity.

A largely unexplored yet critical question is therefore how to intelligently partition amplitude and phase learning between classical and quantum resources. Suppose the amplitude landscape can be captured classically\cite{bluvstein2021controlling,lange2024architectures,pfau2024accurate,hermann2020deep,carleo2017solving} while reserving quantum resources for the inherently hard phase structure\cite{westerhout2023many,park2025complexity,pagano2020quantum,leontica2024exploring,diez2024connection,szabo2020neural,westerhout2020generalization,nomura2017restricted,cai2018approximating,carleo2017solving,preskill2018quantum}. In that case, the quantum circuit may remain shallow, measurement-efficient, and less affected by noise, without sacrificing overall expressivity. However, realizing such a division of labor poses its own challenges. The wavefunction representation must remain valid, optimization must be stable across heterogeneous parameter spaces, and the classical component must not eliminate the need for quantum computation altogether. Unlike classical variational quantum Monte Carlo methods, where amplitude-phase decomposition is straightforward, integrating quantum circuits introduces unique technical hurdles, such as ensuring non-commutative layers do not entangle the separated components and maintaining scalability under hardware constraints.

In this work, we show that such a separation is not only possible but offers clear scaling advantages. We introduce sVQNHE, a variational framework in which a neural network (NN) models the amplitude distribution and a shallow diagonal quantum circuit learns the phase (sign) structure through neural-guided layer-wise optimization. To enable stacking multiple phase layers without collapse, since diagonal two-qubit gates alone support only quadratic interactions, we insert shallow, classically simulatable non-diagonal gates (e.g., a layer of \(R_y\) gates) via a neural-guided amplitude transfer: the NN's learned distribution is iteratively approximated onto these gates through classical simulations, preserving the decomposition by isolating amplitude uploads from phase encoding. This targeted quantum phase learning ensures that (i) dramatically reduced measurement cost due to commuting Pauli terms in the circuit, (ii) a well-conditioned optimization landscape that mitigates barren plateaus, (iii) significantly improved noise resilience from controlled circuit depth and NN mitigation, and (iv) more robust shot-based estimation of observables compared to the other non-unitary hybrid quantum-neural method. Notably, this design does not diminish quantum expressivity—neural amplitude modeling and quantum phase learning are jointly trained to reproduce nontrivial many-body states and combinatorial structure, representing a key innovation that bridges classical deep learning's function approximation power with quantum hardware's interference and entangling capabilities.

We demonstrate sVQNHE's effectiveness across strongly correlated quantum many-body systems and large-scale combinatorial optimization. On the 6-qubit frustrated J1-J2 model, sVQNHE reduces mean absolute energy error by $98.9\%$ and variance by $99.6\%$ versus the classical NN baselines, converging $\sim 19 \times$ faster than hardware-efficient VQE. It also resolves molecular ground states such as the $H_2O$ molecule and demonstrates robustness under depolarizing noise and finite sampling. In the Maximum Clique problem (135 vertices, p=0.5), sVQNHE using only 10 qubits via Pauli-correlation encoding, substantially outperforms greedy heuristics, and a recently proposed Free-Energy Machine \cite{shen2025free} (FEM), thereby confirming the non-trivial contribution of its quantum phase-learning module in highly constrained, rugged landscapes. Furthermore, on a MaxCut instance with 1,485 vertices, it achieves the same performance as the state-of-the-art, classical FEM using only 12 qubits. These results highlight sVQNHE's enhanced generalization across many-body physics and diverse combinatorial optimization challenges, including problems where classical solvers like FEM face limitations due to mean-field approximations. This provides evidence that explicit quantum phase learning, tightly coupled with classical amplitude modeling, constitutes a principled paradigm for scalable hybrid algorithms. By optimizing the co-learning strategy on the amplitude and phase at the representation level, sVQNHE offers a pathway toward reliable near-term quantum utility in quantum chemistry, materials science, and optimization, enabling NISQ-era devices to deliver meaningful performance before the arrival of full fault tolerance.

\section*{Results}
\subsection*{Theoretical Framework of sVQNHE}
\begin{figure*}[htp!]
\centering
\includegraphics[width=0.8\textwidth]{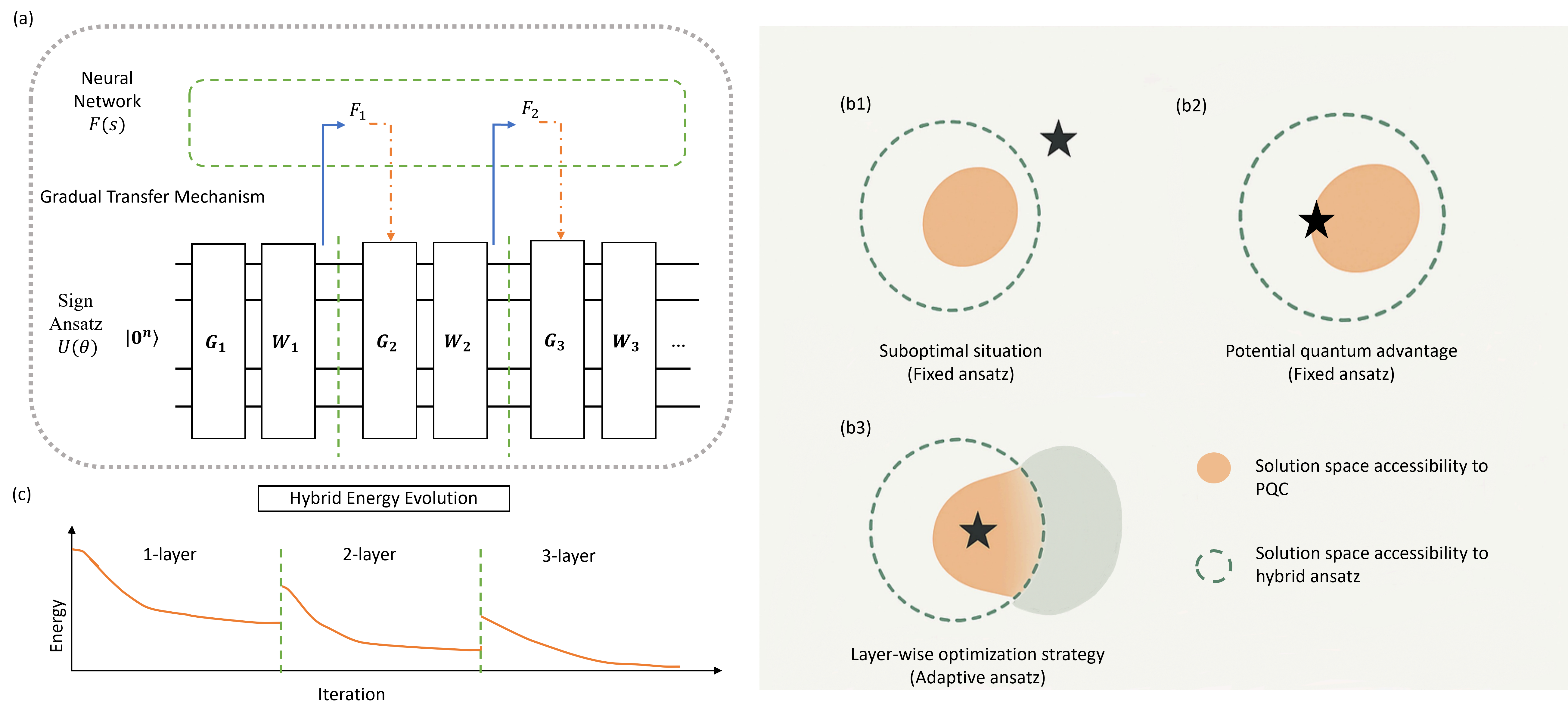}
\caption{(a) The algorithm seamlessly integrates quantum circuits with neural networks. The quantum circuits interleaves diagonal layers ($W_1 \cdots W_n $) with shallow, simultable layers ($G_1 \cdots G_n$). Parameters of the $G$ layers are optimized classically using a gradual transfer mechanism, while those of the $W$ layers are adjusted via the parameter shift rule on the actual hardware. Leveraging the commuting properties of the diagonal layers, the number of measurements needed to compute gradients for the $d$ parameters of $W$ drops from $\mathcal{O}(2d) $ to $ \mathcal{O}(1)$. (b) Illustration of sVQNHE's advantages: The green dashed circle represents the coverage of the hybrid quantum-classical scheme. The star represents the optimal solution. (b1) describes the suboptimal situation, that is, the PQC cannot connect the solution region with the optimal solution;  (b2) describes the potential quantum advantage. Similar to (b1), but the star is now inside the solution space, indicating successful coverage. (b3) Adjustable ansatz shows a transition from (b1) to (b2) via parameter adjustment, including gradient shading to indicate the evolution. (c) Illustration of the energy optimization process using the hierarchical optimization scheme. When entering the next layer, the local optimal solution will jump out due to the newly introduced parameters.} 
\label{fig:framework}  
\end{figure*}

We introduce sVQNHE, a hybrid quantum-classical variational algorithm designed to address key limitations of conventional VQAs, including high measurement cost, barren plateaus, limited expressive power with shallow circuits, and noise-induced errors. This framework builds on the original VQNHE approach by introducing a layer-wise, phase-amplitude decoupling strategy\cite{westerhout2023many} (see Figure \ref{fig:framework}) that enhances trainability, scalability, and sampling robustness. 

To elucidate the algorithm, we begin by summarizing the redesign of the quantum circuit and classical NN ansatzes. The overall (unnormalized) ansatz takes the form: 
\begin{eqnarray}\label{eq:ansatz}
    \ket{\psi}=F_L\prod_{i=1}^{L}W_i G_i \ket{0},
\end{eqnarray}
where $F_L$ denotes a post-processing operator defined by a classical NN. The following paragraphs explain the design of the hybrid ansatz and the associated bidirectional optimization strategy. Finally, the algorithm is recapped in Alg. \ref{alg:sign-vqnhe}.

\textit{Quantum Circuit for Phase Structure Learning.} The quantum circuit ansatz consists of alternating layers of parameterized gates:$\left( G_i, W_i\right)$. $G_i$ are classically simulatable and non-diagonal layers (e.g., $R_y$ gates), primarily responsible for shaping the amplitude structure. $G_1$ is usually taken to be a layer of Hadamard gates to create a uniform superposition, but it can certainly assume an arbitrary form like other $G_i$ layers.   
$W_i$ must be diagonal gates (e.g., $R_z$ and $R_{zz}$) designed to encode the phase structure of the wavefunction. These layers are added iteratively when greater expressivity is needed, and their parameters are optimized in a layer-wise fashion. 

\textit{Classical Neural Network for Amplitude Modeling.} The classical neural network is constrained to produce non-negative outputs, and is used to model the amplitude $F = \sum_s f(s) \ket{s}\bra{s}$ of the quantum state in the computational basis. With the phase information delegated to the quantum circuit, the NN is focused on learning the probability distribution. Previous works have shown that classical neural networks are highly effective in this task, particularly for systems with smooth or structured amplitude landscapes \cite{westerhout2020generalization,westerhout2023many, szabo2020neural}, relevant for many physical applications.

\textit{Bidirectional Feedback Loop: Decoupling Amplitude and Phase.} At each iteration $i$, a new block of parametrized quantum layers $(G_i,W_i)$ is added to the circuit. The process follows two key steps. 
(1) Forward Initialization Step: The classical NN operator $F$ from the previous iteration is approximated by the shallow quantum layer $G_i$, via a tractable classical optimization procedure as detailed in the Method section. This ensures that $G_i$ inherits the amplitude structure previously learned by the NN.
(2) Hybrid Optimization: Once $G_i$ has learn an approximation of $F$ and is loaded onto the quantum circuit, the neural operator $F$ is reset and jointly optimized with the parameters of the new diagonal gate layer $W_i$ by minimizing a loss function, such as the energy of a given Hamiltonian. This hybrid optimization simultaneously refines both the amplitude ($F$) and sign structure ($W_i$) in a tightly coupled quantum–classical feedback loop. 

\begin{algorithm}[!ht]
\scriptsize
\caption{sVQNHE Hybrid Algorithm} 
\label{alg:sign-vqnhe}
\begin{algorithmic}[1]
\Require $n$-qubit system, target Hamiltonian $H = \sum_i c_i P_i$
\Ensure Optimized energy value $E_{\text{final}}$

\Statex \textbf{Initialization}
\State Initialize quantum state $\ket{\psi_0} \gets \ket{0}^{\otimes n}$ 
\State Initialize classical neural network parameters
\State Define diagonal operator $F_1 \gets \sum_s f(s) \ket{s}\bra{s}$ \Comment{$f(s)$: NN-predicted amplitude}
\State Set $G_1 \gets H^{\otimes n}$ \Comment{Uniform superposition}

\Statex \textbf{Joint Optimization}
\While{$\neg \text{converged}$} \Comment{Global convergence}
    \For{$l \gets 1$ \textbf{to} $L$} \Comment{Layer-wise optimization}
        \If{$l > 1$}
            \State Update $\theta^{g_l}$ to minimize $\|G_l - F_{l-1}\|$ \Comment{Gradual Transfer Mechanism, see Method for detail}
        \EndIf
        \State $E_l \gets 0$ \Comment{Reset layer energy}
        
        \Statex \textbf{Hybrid Energy Estimation}
        \Repeat
            \State Apply operator: $\ket{\psi_l} \gets W_l(\theta^{w_l}) G_l(\theta^{g_l}) \ket{\psi_{l-1}}$
            \State Construct hybrid state: $\ket{\psi_{l,f}} \gets F_l \ket{\psi_l}$
            
            \For{\textbf{each} Pauli term $P_i \in H$} 
                \State Compute $E_{P_i} \gets \dfrac{\bra{\psi_{l,f}} P_i \ket{\psi_{l,f}}}{\braket{\psi_{l,f}}{\psi_{l,f}}}$
                \State Accumulate energy: $E_l \gets E_l + c_i \cdot E_{P_i}$ 
            \EndFor

            \State Update NN parameters for $F_l$
            \State Update parameters: $\theta^{w_l}$ \Comment{Measurement Strategy, see Method for detail}
        \Until{$|\Delta E_l| < \epsilon_{\text{conv}}$} \Comment{Energy convergence} 
    \EndFor
    \State Check global convergence \Comment{Track $E_l$ across layers}
\EndWhile
\State \Return $E_L$ \Comment{Final optimized energy}
\end{algorithmic}
\end{algorithm}

\noindent\textbf{\\ Efficient Allocation of Quantum Resources}

Traditional variational algorithms, such as the Variational Quantum Eigensolver (VQE) and the Quantum Approximate Optimization Algorithm (QAOA), typically employ fixed circuit ansatzes derived from physical intuitions like Trotterized time evolution\cite{aaronson2009need,mcclean2021low}. These static architectures can limit performance when the ansatz fails to capture the entanglement or correlations required for the target state, often trapping optimization in local minima. Adaptive strategies like adapt-VQE attempt to address this by iteratively selecting gates from a predefined pool, but they remain constrained by the expressiveness of the pool and incur high measurement overhead due to repeated gate evaluations\cite{tang2021qubit,grant2019initialization}.

In contrast, sVQNHE employs a neural-guided adaptive hybrid ansatz (as defined in Equation \eqref{eq:ansatz}) that achieves an alternative and (potentially) more efficient allocation of quantum resources. The ansatz alternates between shallow and non-diagonal $G_i$ (e.g., $R_y$ rotations) and diagonal multi-qubit phase-encoding layers $W_i$ (e.g., $R_z$ and $R_{zz}$). Crucially, circuit growth proceeds in a layer-wise manner under the guidance of the classical neural post-processing module $F$. At each stage of optimization, the neural network receives the current output distribution from the quantum circuit and provides feedback that informs whether additional layers are required and, if so, how they should be structured to most effectively refine the approximation. This amplitude-guided, progressive deepening ensures that quantum resources are deployed selectively and only where necessary to capture residual discrepancies in the target distribution, avoiding unnecessary depth or parameter overhead.

Complementing this adaptive construction is a measurement strategy that exploits the commuting structure within each layer: the diagonal $W_i$ layers are fully commuting with extremely low measurement costs, while the parameters in shallow, simulatable $G_i$ layers are optimized classically without any measurement costs. As a result, gradient estimation requires substantially fewer quantum shots compared with the standard VQE algorithm employing the hardware-efficient or other general, entangling ansatzes, thereby reducing the overall measurement burden and enhancing scalability on devices with limited shot capacity. A detailed discussion of the measurement-efficient gradient protocol is provided in the Measurement Reduction part.

Even with relatively simple non-diagonal layers based on $R_y$ rotations, the chosen generator set $\{Z D_{2j}, Y_i\}$ spans a Lie algebra of dimension $\mathfrak{su}(2^{n-1}) \oplus \mathfrak{su}(2^{n-1})$ (Theorem \ref{the:lie_ansatz}), enabling the approximation of highly entangled states using only shallow circuit depth. Furthermore, the incorporation of neural post-processing helps maintain polynomial gradient variance in the loss landscape, $\mathrm{Var}[\partial_{\theta_k} \mathcal{L}] \in \Omega(1/\mathrm{poly}(n))$, thereby suppressing the emergence of barren plateaus during adaptive growth (Theorem \ref{the:landscape}). Proposition \ref{propLayer} additionally informs the design of non-commuting layer sequences, ensuring that added depth contributes meaningfully to expressiveness.

Consequently, sVQNHE dynamically assembles shallow yet highly expressive circuits whose reachable Hilbert space expands in a targeted, data-driven fashion until the ground-state wavefunction or desired combinatorial solution distribution is adequately represented (illustrated schematically in Figure \ref{fig:framework}(b3)). Empirical results across frustrated spin models (such as J1–J2 chains) and large-scale combinatorial optimization instances (including MaxCut and Maximum Clique under qubit-efficient encodings) demonstrate that sVQNHE consistently achieves high-fidelity output distributions using fewer quantum layers and markedly lower measurement costs than fixed-depth variational baselines, non-adaptive hybrid methods, or purely classical neural approaches, even for distributions that are classically difficult to sample due to strong entanglement and complex sign structure \cite{troyer2005computational}.

In summary, by integrating neural-directed adaptive growth, measurement-frugal layer designs, and Lie-algebra-enhanced shallow-circuit expressiveness, sVQNHE reallocates quantum resources far more efficiently than traditional fixed or purely adaptive variational frameworks. This resource-aware construction positions the method as a robust candidate for realizing practical quantum utility on near-term noisy quantum devices, where both qubit count and measurement budget remain severely constrained.

\noindent\textbf{\\ Reducing Measurement Costs and Enhancing Trainability}

Traditional variational quantum algorithms face two intertwined challenges that limit scalability: high measurement costs and the emergence of barren plateaus. The first challenge arises from the need to estimate gradients by measuring non-commutative operators, a process requiring $\mathcal{O}(2d)$ measurements for $d$ trainable parameters. In large systems with $\text{poly}(n)$ operators, this results in a prohibitive $\mathcal{O}(2d \cdot \text{poly}(n))$ measurement overhead, which becomes computationally intractable as circuits scale. The second challenge, barren plateaus, manifests as exponentially vanishing gradients in high-dimensional parameter spaces, rendering optimization ineffective. These issues are exacerbated in deep circuits and systems with many qubits, where the parameter landscape grows increasingly complex and gradients vanish rapidly due to the curse of dimensionality.  

The sVQNHE framework addresses these limitations through two synergistic innovations. First, it employs commuting diagonal gates, such as $R_z$ and $R_{zz}$ rotations, which enable simultaneous measurement of all relevant expectation values in one basis. This compresses the per-operator measurement cost to $\mathcal{O}(1)$, reducing the total overhead to $\mathcal{O}(\text{poly}(n))$ for $\text{poly}(n)$ operators (see details in the method section). This stands in stark contrast to the multiplicative $\mathcal{O}(2d \cdot \text{poly}(n))$ scaling of traditional VQNHE and conventional VQAs, allowing sVQNHE to tackle large-scale problems efficiently. 

Secondly, our framework incorporates layer-wise neural guidance to decompose the optimization process into manageable subproblems, addressing a critical limitation of traditional layer-wise VQAs\cite{liu2022layer}. In conventional layer-wise VQAs, the absence of a global perspective on the optimization landscape renders sequential layer optimization prone to converging at local minima. In contrast, the sVQNHE framework employs a classical neural network to guide the optimization of each quantum layer, producing a hybrid state $|\psi_{l,f}\rangle$ that is more expressive than the state $|\psi_l\rangle$ obtained from hierarchical quantum circuits alone. This neural guidance not only helps avoid suboptimal local minima but also focuses parameter updates on one specific block each time, such as $W_l$ at the $l$-th iteration. This approach, supported by the benign landscape of Theorem \ref{the:landscape}, restricts the effective parameter space per iteration (e.g., block \(W_l\)), preventing convergence to local minima and maintaining scalability. Together, these strategies circumvent the dual bottlenecks of measurement overhead and barren plateaus, establishing a scalable pathway for variational quantum algorithms in practical applications.

\noindent\textbf{\\ Sampling Noise Robustness}

Under the finite sampling conditions, the mixture energy estimate in VQNHE requires normalization to counteract the non-unitary transformation induced by the neural network module $F$. The variance of the energy estimate $\hat{E}_{P_i}$ is governed by:
\begin{align}\label{eq:var}
    \text{Var}(\hat{E}_{P_i}) = &\text{Var}\left(\bra{\psi_{L,f}} P_i \ket{\psi_{L,f}}\right) \cdot \text{Var}(g^{-1}(s,L)) + \text{Var}\left(\bra{\psi_{L,f}} P_i \ket{\psi_{L,f}}\right) \cdot\mathbb{E}^2(g^{-1}(s,L)) \nonumber\\
    &+ \text{Var}(g^{-1}(s,L))\cdot \mathbb{E}^2\left(\bra{\psi_{L,f}} P_i \ket{\psi_{L,f}}\right) + \mathcal{O}(\text{Cov}(\bra{\psi_{L,f}} P_i \ket{\psi_{L,f}},g^{-1}(s,L)),
\end{align}
where $g(s,L) = \frac{1}{n_s}\sum_{s_i} f^2(s_i)$, and $s_1, \ldots, s_{n_s}$ are i.i.d. samples drawn from the distribution $|\langle s | \psi_L \rangle|^2$. The expectation and variance of $g(s,L)$ are $\mathbb{E}[g(s,L)] = \mathbb{E}\left[f^2(s)\right]$ and $\text{Var}(g(s,L)) = \frac{1}{n_s} \text{Var}\left[f^2(s)\right]$, where $s \sim |\langle s | \psi_L \rangle|^2$. This normalization process amplifies the sampling noise in the mixture energy estimate compared to the standard VQE, as quantified by $\text{Var} (g^{-1}(s,L)) > 1$. Notably, when the number of layer $L = 1$ is fixed, this formulation form simplifies to that of the original VQNHE algorithm. 

The sVQNHE algorithm introduces a layer-adaptive gradual transmission mechanism. As the quantum circuit depth $L$ increases, the algorithm shifts computational responsibility to the quantum component, progressively diminishing the weighting effect of $f(s)$ across configurations $s$. This mechanism ensures: 
\begin{equation}
    \text{Var}_{s \sim |\langle s | \psi_L \rangle|^2}\left(f^2(s)\right) \xrightarrow{L \uparrow} 0.
\end{equation}
For sufficiently large $L$, the variance of $f^2(s)$ becomes small ($\text{Var}(f^2(s)) < 1$), which implies that $g(s,L)$ concentrates tightly around its mean $\mathbb{E}[g(s,L)] = \mathbb{E}\left[f^2(s)\right]$.

Under this concentration condition, the variance of the normalization factor $g^{-1}(s,L)$ can be accurately approximated using the first-order delta method (also known as the linearization method):
\begin{equation}
   \text{Var}\bigl( g^{-1}(s,L) \bigr) \approx \frac{1}{\mathbb{E}^4[g(s,L)]}\text{Var}(g(s,L)) = \frac{1}{\mathbb{E}^4[g(s,L)]} \text{Var}_{s \sim |\langle s | \psi_L \rangle|^2}\bigl[ f^2(s) \bigr].
\end{equation}
As $L \to \infty$, the numerator $\text{Var}(f^2(s))$ tends to zero while the denominator remains bounded, implying $\text{Var}(g^{-1}(s,L)) \to 0$. Moreover, the covariance term $\text{Cov}\left(\bra{\psi_{L,f}} P_i \ket{\psi_{L,f}}, g^{-1}(s,L)\right)$ also diminishes with increasing circuit depth $L$. This is because as $L$ increases, the quantum state $\ket{\psi_{L,f}}$ progressively approaches the true ground state, causing the Pauli expectations $\bra{\psi_{L,f}} P_i \ket{\psi_{L,f}}$ to stabilize, while $\text{Var}(g^{-1}(s,L))$ simultaneously vanishes as established above. The synchronous reduction in fluctuations of both variables naturally leads to a decaying covariance. Consequently, for sufficiently large $L$, the covariance term becomes negligible, and the variance formula is dominated by the first three terms in equation \ref{eq:var}. As a result, all contributions involving $\text{Var}(g^{-1}(s,L))$ or the covariance vanish asymptotically, and $\text{Var}(\hat{E}_{P_i})$ converges to the variance level of the standard VQE, indicating that the sampling noise amplification effect in original VQNHE is mitigated in the deep-circuit regime.

In our numerical experiments, as illustrated in Figures~\ref{fig:merged_sampling}(a) and (b), the coefficient of variation of the energy estimate $\text{CV}(\hat{E}) \propto \sqrt{\text{Var}(g^{-1}(s,L))}$ decays with increasing $L$. This confirms that sVQNHE exhibits improved noise immunity and robustness under realistic finite-sampling constraints as the quantum circuit depth grows.
In summary, we mitigate a source of numerical instability of the original VQNHE.

\subsection*{Experimental Results}
To systematically evaluate the proposed sVQNHE framework, we design a 4-stage benchmark suite that progressively demonstrates its key advantages: (1) its ability to overcome optimization challenges arising from complex phase structures in strongly frustrated quantum many-body systems, (2) its general performance improvement even in systems with trivial phase structures due to enhanced optimization landscapes and measurement efficiency, (3) its robustness and noise mitigation capabilities under real NISQ conditions (without any error mitigation), and (4) its potential to deliver near-term quantum utility in large-scale combinatorial optimization using a qubit-efficient encoding.

All quantum simulations are performed without any form of quantum error mitigation. Noisy results are obtained directly from shot-based sampling of noisy quantum circuits. Unless otherwise stated, energy values are reported as relative errors with respect to the exact ground-state energy, and all results are averaged over multiple random initializations.

\subsubsection*{1. Frustrated Quantum Many-Body Systems}
The experimental application of the sVQNHE framework to frustrated quantum spin systems serves as a critical validation of its advantages. These systems, characterized by highly complex ground-state phase structures and strong correlations, pose significant challenges to conventional variational quantum algorithms. Because it requires deep circuits to optimize both amplitude and phase components simultaneously, they often encounter optimization challenges, such as getting stuck in barren plateaus or local minima \cite{mcclean2018barren,cao2025exploiting}. By explicitly separating these tasks, delegating phase learning to a shallow quantum module while modeling the amplitude with classical neural networks, sVQNHE effectively mitigates the complex phase structures problem. The resulting faster and more stable convergence demonstrates that sVQNHE successfully alleviates key optimization dilemmas inherent to traditional VQEs, thereby achieving superior performance in strongly correlated systems compared to purely classical or purely quantum approaches and highlighting the framework’s potential as an efficient hybrid strategy for tackling complex quantum many-body problems.

We focus on two paradigmatic models with highly non-trivial ground-state phase diagrams and strong sign oscillations in the computational basis to test:
\begin{itemize}
    \item[-]  the 1D J1–J2 chain with $J_1/J_2 = 0.6$ (in the Majumdar–Ghosh regime near the dimerized phase, known for its challenging phase structure \cite{viteritti2022accuracy}),
    \item[-] the 2D square-lattice Heisenberg antiferromagnet with periodic boundary conditions (9 qubits, exhibiting severe phase complexity in numerical methods \cite{stoudenmire2012studying}).
\end{itemize}
We compare sVQNHE (sign-VQNHE) against four strong baselines:
\begin{itemize}
    \item[-] pure classical neural network (NN) with real-valued MLP,
    \item[-] hardware-efficient VQE (HEA) with 1 or 2 layers,
    \item[-] original VQNHE using the same hardware-efficient quantum circuit and real-valued NN (denoted hea-NN),
    \item[-] QAOA with up to 10 layers.
\end{itemize}
All methods are trained with the same number of optimization steps (2000) and evaluated over 20 random initializations. Performance is quantified using relative mean absolute error ($R_{\text{MAE}}$) and relative variance ($R_{\text{Var}}$) with respect to the pure NN baseline (see Methods for definitions; values closer to -1 indicate better performance). Results are summarized in Table \ref{tab:all_model}.

\begin{table}[htp!]
\centering
\scalebox{0.6}{
\begin{tabular}{@{}cccccccccccccccc@{}}
\toprule
\multirow{2}{*}{\textbf{Metric}} & \multicolumn{3}{c|}{\textbf{6q J1-J2}} & \multicolumn{3}{c|}{\textbf{9q Heisenberg}} & \multicolumn{2}{c|}{\textbf{9q TFIM}} & \multicolumn{4}{c|}{\textbf{9q Ising}} & \multicolumn{3}{c}{\textbf{12q Ising}} \\ \cmidrule(l){2-16} 
 & sign-VQNHE & hea-NN & HEA2 & HEA1 & hea-NN & sign-VQNHE & HEA1 & sign-VQNHE & HEA1 & QAOA5 & QAOA10 & sign-VQNHE & HEA1 & QAOA10 & sign-VQNHE \\ \cmidrule(r){1-1}
$R_{\text{MAE}}$ & \textbf{-0.989} & -0.751 & -0.916 & -0.331 & -0.119 & \textbf{-0.872} & -0.290 & \textbf{-0.640} & 0.129 & 20.674 & 1.205 & \textbf{-0.416} & 6.822 & 77.012 & \textbf{-0.811} \\
$R_{\text{Var}}$ & \textbf{-0.996} & 1.955 & -0.132 & 0.554 & 33.793 & \textbf{-0.832} & 0.019 & \textbf{-0.728} & 0.909 & 15.583 & 11.835 & \textbf{-0.455} & 10.002 & 80.404 & \textbf{-0.979} \\ \bottomrule
\end{tabular}}
\caption{\small Performance Analysis of sVQNHE Algorithm. Here, the range of $R_{MAE}$ and $R_{Var}$ is $(-1,+\infty)$. The closer to $-1$, the better (see Equations \eqref{eq:R_MAE} and \eqref{eq:R_var} in the Method section for details). The baseline algorithm NN model is a real Multi-Layer Perceptron (MLP). HEA1 and HEA2 represent VQE methods employing 1-layer and 2-layer Hardware-Efficient Ansatz, respectively. QAOA5 and QAOA10 represent the QAOA method with layers 5 and 10, respectively. The Sign Ansatz employs $R_{zz}$ gates to connect all edges of the corresponding physical model.}
\label{tab:all_model}
\end{table}

For the 6-qubit J1-J2 chain, sVQNHE achieves dramatic improvements. Specifically, $R_{\text{MAE}} = -0.989$ implies $98.9\%$ error reduction vs. NN, and $R_{\text{Var}} = -0.996$ implies $99.6\%$ variance reduction vs. NN. In contrast, hea-NN only reduces MAE by $75.1\%$ while increasing variance compared to NN. Conventional 2-layer HEA (HEA2) reaches -0.916 in MAE but still lags far behind sVQNHE in both accuracy and stability. Moreover, sVQNHE converges significantly faster: it requires on average $12.6\times$ fewer steps than HEA2 to reach the same error threshold, and succeeds in reaching target accuracy $12\times$ more frequently. In the most efficient cases, HEA2 needed 1676 steps to reach the same accuracy, while sVQNHE only required 89 steps (for details, please refer to the supplementary material). Furthermore, due to the non-commutative structure of hardware-efficient ansatze, the measurement overhead for sVQNHE is significantly lower than for hea-NN under similar parameter counts.

\begin{figure}[htp!]
\centering
\includegraphics[width=0.48\textwidth]{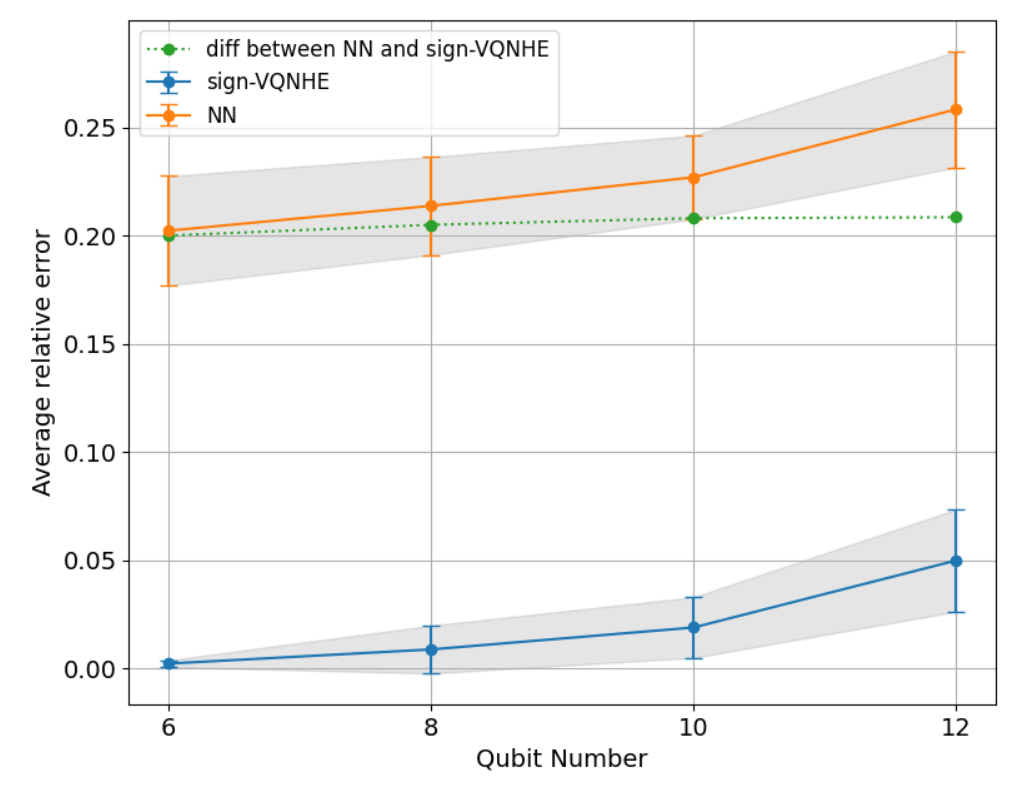}
\caption{Scaling test for the 1D J1-J2 model. The x-axis shows the number of qubits, while the y-axis displays the average relative error from the sVQNHE method (sign-VQNHE, blue line) and the NN method (NN, orange line) after 20 runs of 2000 steps each. The green line illustrates the difference in average relative error between the NN and sign-VQNHE methods as the qubit count increases. Shaded regions represent the relative error bar scaled by 0.5.} 
\label{fig:J1-J2}  
\end{figure}

The scaling behavior is even more striking. Figure \ref{fig:J1-J2} shows the performance gap between sVQNHE and NN as system size increases from 6 to 12 qubits under a fixed optimization budget. The advantage of sVQNHE widens systematically with system size, demonstrating that the phase-focused architecture becomes increasingly valuable as correlation strength and phase complexity intensify.

Similar qualitative behavior is observed on the 9-qubit 2D Heisenberg model ($R_{\text{MAE}} = -0.872$, $R_{\text{Var}} = -0.832$), confirming that sVQNHE consistently resolves the trainability barriers faced by both pure classical NN (which cannot model phases) and hardware-efficient variational circuits (which entangle amplitude and phase learning) in regimes with complex phase structures.

\subsubsection*{2. Trivial Phase Structures Systems}
Having demonstrated sVQNHE’s strength in systems with complex phase structures, we next ask whether the method remains advantageous in systems where the ground-state phase is trivial (i.e., real-valued or easily captured without specialized phase modeling).

We consider three representative cases without significant phase frustration:
\begin{itemize}
    \item[-] 1D Transverse Field Ising Model (TFIM, 9 qubits, exactly solvable via Jordan–Wigner transformation),
    \item[-] 1D classical Ising chains (9 and 12 qubits, relevant to combinatorial optimization encodings),
    \item[-] electronic ground state of the water molecule ($H_2O$) in STO-3G basis (10 qubits after active-space reduction, introducing electronic correlation challenges).
\end{itemize}

Table \ref{tab:all_model} shows that even without complex phases, sVQNHE systematically outperforms the pure NN baseline. Notably, the performance margin increases with system size, indicating improved scalability. When benchmarked against QAOA (5 or 10 layers), sVQNHE maintains its edge: QAOA requires more measurements per iteration and yields higher errors, even at deeper layers. The superior performance stems from sVQNHE's phase-focusing architecture (using $R_{zz}$ and $R_y$ gates with independent parameters), which generates a higher-dimensional Lie algebra compared to QAOA (as per Theorem \ref{the:lie_ansatz}), enabling broader exploration of the unitary space.

\begin{figure}[htp!]
\centering
\includegraphics[width=0.7\textwidth]{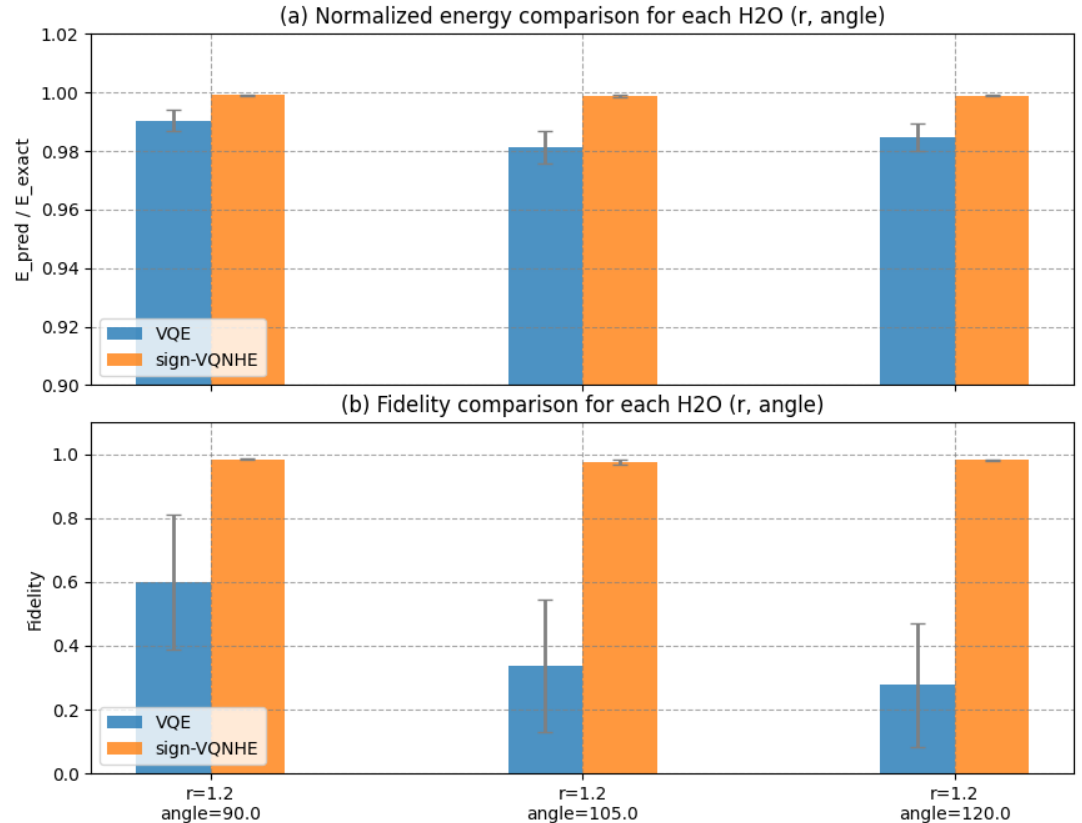}
\caption{Comparison of normalized energy and fidelity for $H_2O$ molecules at various bond angles (angle). The x-axis labels denote different geometric configurations (r, angle), where (r) is the bond length. The top panel (a) presents the average normalized energy \( E_{\mathrm{pred}} / E_{\mathrm{exact}} \) for each group, while the bottom panel (b) shows the average fidelity between the predicted and exact wavefunctions. Blue bars represent results from the VQE method, and orange bars correspond to the sVQNHE method. Error bars indicate 0.5 standard deviation over 5 independent runs. } 
\label{fig:H2O}  
\end{figure}

For the $H_2O$ molecule, we compare sVQNHE against standard 3-layer hardware-efficient VQE across multiple geometries (generated with \texttt{PySCF}, bond length $1.2$, bond angles $[90^\circ,120^\circ]$). Both methods use comparable total quantum parameter counts. As shown in Figure \ref{fig:H2O}, sVQNHE consistently achieves: lower normalized energy error and higher wavefunction fidelity. Across the entire geometry scan, despite using half as many quantum parameters per iteration and benefiting from fully-commuting layers (leading to dramatically lower measurement cost).

These results demonstrate that the division of the amplitude–phase task between quantum and classical resources with neural-guided optimization significantly improves trainability and solution quality even when the phase learning is not particularly challenging. The benefits arise from a smoother optimization landscape, reduced variance, and enhanced measurement efficiency, advantages that are broadly relevant across many problem classes.

\subsubsection*{3. Robustness Under Noisy Conditions}
To assess the practical viability of sVQNHE in the NISQ regime, we evaluate its performance under two realistic sources of imperfection: (1) finite sampling shot noise from bitstring measurements, and (2) hardware gate and readout noise (modeled via depolarizing channels). All results are obtained directly from noisy circuit sampling without applying any quantum error mitigation techniques.

We focus on the 3-qubit 1D Heisenberg Hamiltonian as a representative frustrated system. We compare sVQNHE against two VQE variants using the identical Sign Ansatz: layered optimization VQE (layerwise-VQE) and standard optimization VQE (standard-VQE). sVQNHE employs non-negative real-valued neural networks for amplitude correction, while both VQE baselines optimize only quantum parameters. All simulations are capped at 400 iterations, with resource-efficient commuting-group measurements for VQE.

\begin{figure}[htp!]
\centering
\begin{subfigure}[b]{0.4\textwidth}
    \includegraphics[width=\textwidth]{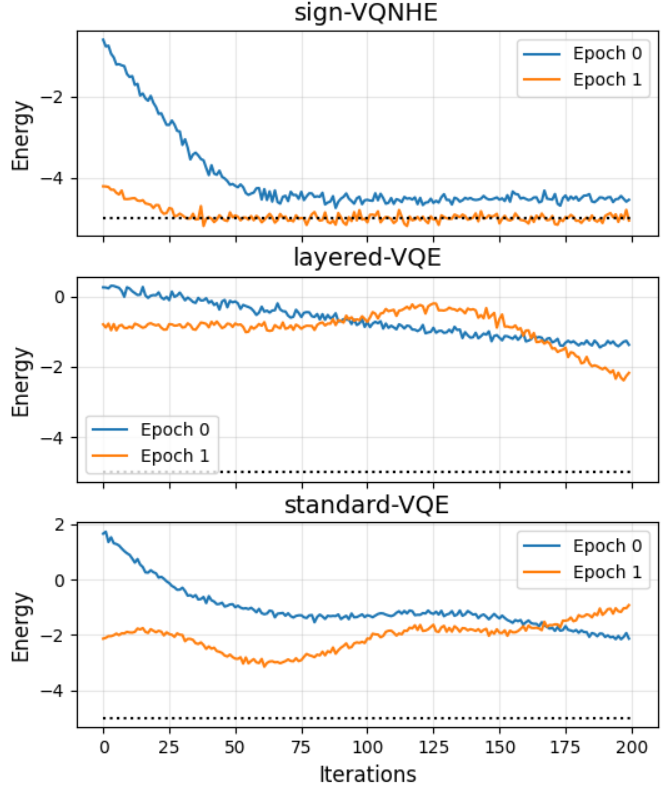}
    \caption{Performance evaluation under finite sampling.}
    \label{fig:sampling} 
\end{subfigure}
\hspace{0.02\textwidth}
\begin{subfigure}[b]{0.403\textwidth}
    \includegraphics[width=\textwidth]{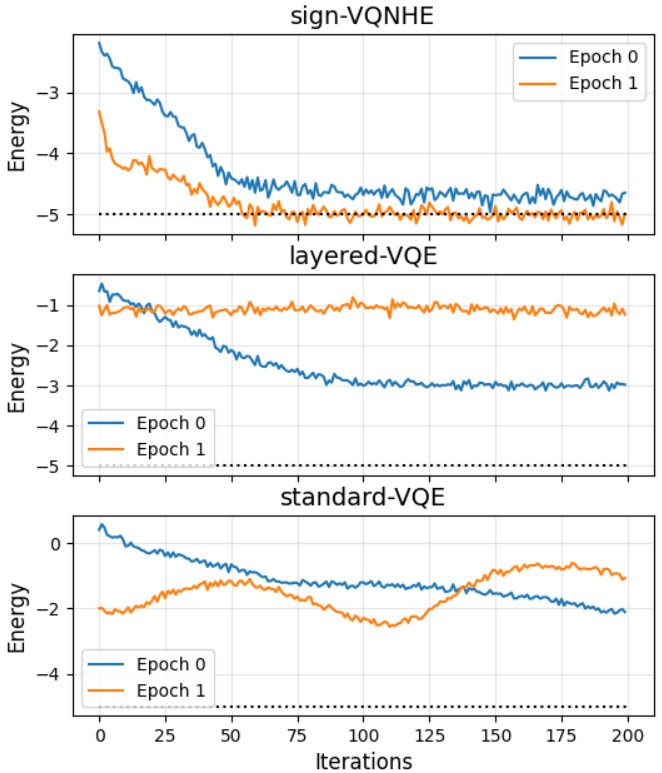}
    \caption{Performance evaluation under noisy conditions.}
    \label{fig:sampling_noise}
\end{subfigure}
\caption{Comparative performance of sVQNHE and VQE variants under a 3-qubit 1D Heisenberg Hamiltonian. (a) Finite sampling (1200 shots); (b) Noisy conditions (1200 shots, two-qubit: $0.005$; single-qubit: $0.0013$). Iteration count (x-axis) versus energy (y-axis). The solid black dashed line marks the ground state energy value. For layered optimization (sVQNHE, layered-VQE), blue and orange lines denote first- and second-layer parameter updates, respectively; standard-VQE (overall optimization) reflects sequential parameter refinement. For sVQNHE, coefficients of variation after stabilization are $cv_1 = 0.01618, cv_2 = 0.015912$ (a) and $cv_1 = 0.01534, cv_2 = 0.01487$ (b), indicating reduced variability in the second layer.}
\label{fig:merged_sampling}
\end{figure}

Figure \ref{fig:merged_sampling}(a) shows convergence curves (energy vs. iteration) under finite sampling (ideal noise-free circuits, limited shots). sVQNHE converges faster and more stably than both VQE variants. Notably, layered optimization fails for pure VQE because the 1-layer Sign Ansatz leads to premature trapping in poor local minima. In contrast, sVQNHE successfully leverages layered growth by delegating initial amplitude modeling to the NN, enabling effective progressive refinement. Post-convergence, the coefficient of variation (CV) in the second layer is lower than in the first ($CV_1=0.01618 \to CV_2 = 0.015912$), confirming that the gradual transfer mechanism enhances stability under sampling variance.

Under realistic device noise (Figure \ref{fig:merged_sampling}(b)), sVQNHE again achieves faster convergence to near-ground-state energies and exhibits lower variability in the refined second layer ($CV_1=0.01534 \to CV_2=0.01487$). This demonstrates that the hybrid quantum-neural architecture possesses an intrinsic error-mitigating capability. The classical NN module can be continuously tuned during noisy optimization to partially compensate for circuit-induced errors, as previously analyzed in related variational hybrid frameworks \cite{zhang2023variational}.

These noisy experiments highlight that sVQNHE not only maintains its advantages in ideal settings but also exhibits superior robustness and reliability under the combined effects of shot noise and incoherent errors.

\subsubsection*{4. Toward Near-Term Quantum Utility in Large-Scale Combinatorial Optimization}
Finally, we investigate whether the sVQNHE framework can deliver meaningful quantum utility in practically relevant combinatorial optimization problems using highly compressed qubit encodings.

We consider two canonical NP-hard problems:
\begin{itemize}
    \item[-] MaxCut Problem (Erdős–Rényi graphs, up to 1485 vertices),
    \item[-] Maximum Clique Problem (Erdős–Rényi graphs, 135 vertices).
\end{itemize}
Both problems are encoded using Pauli Correlation Encodings (PCE) \cite{sciorilli2025towards}, achieving polynomial qubit compression: $O(n^k)$ classical variables are mapped to only $n$ qubits with $k=2$ (MaxClique) or $k=4$ (large MaxCut).

\begin{figure*}[htp!]
\centering
\includegraphics[width=0.65\textwidth]{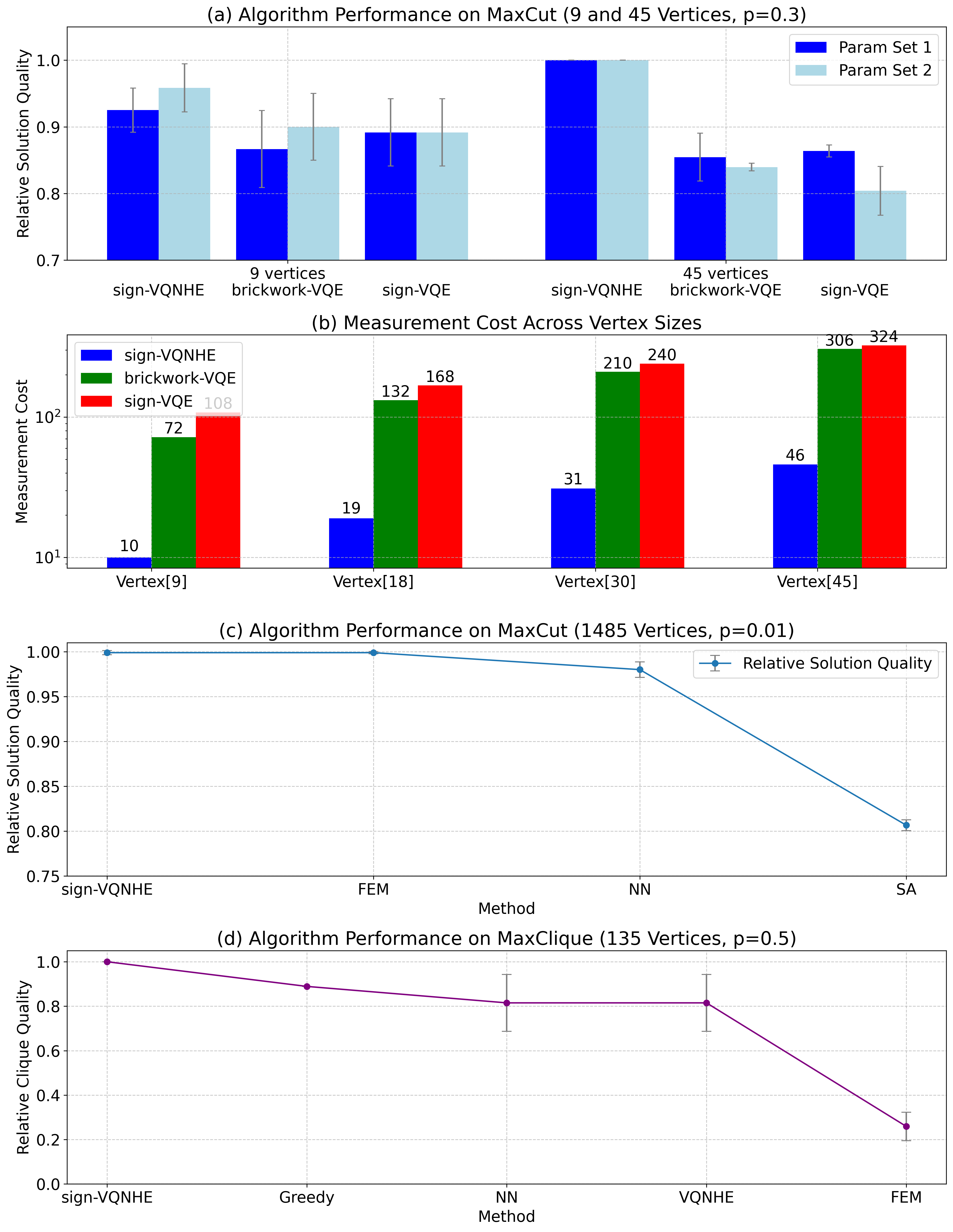}
\caption{Scaling test of sVQNHE using the MaxCut problem as an example. (a) The relative solution quality for the MaxCut problem, comparing parameter sets 1 and 2 across different problem sizes. The methods sign-VQNHE, brickwork-VQE, and sign-VQE correspond to the sVQNHE method with a 2-layer sign ansatz, the VQE method with brickwork ansatz, and the VQE method with 2-layer sign ansatz, respectively. The depth of brickwork ansatz is determined according to the reference \cite{sciorilli2025towards}, that is, the circuit depth is sublinearly related to the number of vertices in the graph. For sign-VQE, the ansatz is the same as the circuit finally generated by the corresponding sVQNHE scheme finally generated, utilizing a two-layer sign ansatz with two diagonal and two non-diagonal layers. The value range of relative solution quality is $[0,1]$. The closer to 1, the better the performance is compared with other algorithms (see Equation \eqref{eq:R_e} in the Method section for details). For example, in the case of 45 vertices, the quality of the solution of sVQNHE is relatively optimal regardless of whether the parameter set is 1 or 2. (b) The single-step measurement consumption for the same problem sizes. Parameter set 2 features an increased number of parameters in the NN component and doubles the number of iterations compared to parameter set 1. The circuit depth in the brickwork architecture scales sub-linearly with the number of vertices. (c) Exact state-vector simulation for MaxCut on Erdős–Rényi graphs with $m=1485$ vertices. Gray error bars represent one standard deviation. (d) Exact state-vector simulation for MaxClique on Erdős–Rényi graphs with $m=135$ vertices. Gray error bars represent one standard deviation.} 
\label{fig:sampling_scaling}  
\end{figure*}

For MaxCut, under noisy sampling conditions, as shown in Figure \ref{fig:sampling_scaling} (a), sVQNHE consistently outperforms the baselines in solution quality. The performance gap widens with scale: for $m=9$, sVQNHE achieves $\sim6\%$ improvement over baselines (e.g., 0.96 vs. 0.90), rising to $\sim19\%$ for $m=45$ (1.00 vs. 0.84) using optimized parameter set 2 (enhanced NN depth and learning rate). Without NN components, sign-VQE and brickwork-VQE perform comparably, indicating that the sign ansatz alone offers no inherent edge over brickwork; the gains stem from sVQNHE's hybrid integration. Figure \ref{fig:sampling_scaling} (b) highlights resource efficiency: sVQNHE reduces single-step measurement costs by $\sim86\%$ for $m=9$ (10 vs. 72) and $\sim85\%$ for $m=45$ (46 vs. 306) compared to baselines. This efficiency arises from NN-guided layer-wise optimization, which minimizes the number of gradient evaluations. Preliminary results in Supplementary Information extend this to larger instances ($m>45$), confirming sVQNHE's scalability under noisy sampling.

For large-scale instances, we use state-vector simulations on graphs with $m=1485$ vertices and edge probability $0.01$, encoded via quartic PCE ($k=4$), requiring $n=12$ qubits. This setup tests the case where direct encodings are infeasible in the NISQ and early FTQC eras. We generate five random instances and compare sVQNHE against: (1) Free-Energy Machine \cite{shen2025free} (FEM), a physics-inspired classical solver minimizing variational free energies via gradient descent; (2) D-Wave's Simulated Annealing Sampler \cite{dwave_ocean_sdk} (SA); and (3) a pure NN solver, using only the classical components of the corresponding sVQNHE method, without the quantum part. For each instance, as in the sampling case, we compute the cut value for all methods and normalize relative to the maximum cut across the four solvers. Averaging over the five graphs yields average relative solution qualities: sVQNHE (0.9989), FEM (0.9989), NN (0.9800), and SA (0.8066) (see Figure \ref{fig:sampling_scaling} (c)). sVQNHE and FEM deliver near-optimal solutions, with sVQNHE slightly outperforming FEM by exploiting quantum correlations for finer exploration of the rugged landscape. The pure NN lags due to underparameterization in high-dimensional spaces, while SA struggles with sparse graphs (low edge probability), as annealing heuristics favor denser structures \cite{johnson1989optimization, myklebust2015solving, hamerly2019experimental}. 

For MaxClique, we evaluate on graphs with $m=135$ vertices and edge probability $p=0.5$. Encoding uses PCE with $k = 2$, compressing to $10$ qubits. Methods include sVQNHE, its pure real NN component (identical classical part without quantum module), VQNHE (hardware-efficient ansatz with equivalent number of quantum parameters), greedy (degree-based iterative addition), and FEM. Non-greedy methods require post-processing (thresholding + greedy repair) to ensure valid cliques, as continuous relaxations may yield infeasible assignments. Over 3 instances, average relative solution qualities, normalized to the maximum across methods) rank as: sign-VQNHE ($1.00$), Greedy ($0.89$), NN and VQNHE ($0.81$), FEM ($0.26$) (see Figure \ref{fig:sampling_scaling} (d)). sVQNHE outperforms greedy's typical $maxclique \approx \log_2 135$ (though greedy requires no post-processing, its local search limits global exploration). The superior performance of sVQNHE highlights the critical value of its quantum module. Compared to the pure NN baseline (which uses the exact same classical neural network component but lacks the quantum phase-learning layers), sVQNHE delivers significantly higher solution quality, demonstrating that the quantum part (diagonal gates $W_i$ for phase structure learning) actively contributes meaningful computational power rather than merely consuming resources. This advantage is further validated against VQNHE (with equivalent parameter count): sVQNHE's innovative co-learning procedure of phase and amplitude components with adaptive layer-wise growth harnesses quantum resources more effectively, avoiding barren plateaus and reducing measurement costs. In addition, although methods relying on continuous energy landscape optimization face greater challenges in this rugged landscape, relative to the continuous optimization method FEM, which claims generality for COPs without Ising reductions \cite{shen2025free}, sVQNHE offers superior global search in constraint-intensive, rugged landscapes. In contrast, FEM's mean-field independence assumption ($\prod_i P_i(\sigma_i)$) fails to capture the strong all-pair connectivity constraints inherent to cliques, leading to diffuse marginal distributions and higher post-processing losses \cite{jagota1995approximating,shen2025free}. This underscores sVQNHE's enhanced universality for combinatorial optimization problems.

These scaling experiments on MaxCut and Maximum Clique demonstrate that sVQNHE consistently delivers superior or competitive performance across diverse combinatorial optimization landscapes. On MaxCut, a problem with relatively smoother energy surfaces, sVQNHE achieves near-optimal solutions comparable to state-of-the-art classical solvers like FEM while offering slight advantages in rugged exploration via quantum correlations, together with substantial reductions in measurement overhead through NN-guided optimization. On Maximum Clique, a paradigmatically rugged, constraint-intensive problem with strong all-pair correlations, sVQNHE markedly outperforms all baselines and validates the critical value of its quantum module. Overall, sVQNHE emerges as a robust, scalable hybrid framework that leverages qubit-efficient encodings and synergistic quantum-classical computation to address both smooth and rugged combinatorial optimization problems effectively, positioning it as a promising candidate for near-term quantum utility in combinatorial optimization.

\section*{Discussion}
The results presented in this work support a broader computational principle: hybrid quantum–classical wavefunction learning becomes significantly more scalable when amplitude and phase are trained by different resources rather than by a single quantum circuit. Recent efforts in variational quantum computing have focused on improving expressivity, ansatz structure, or optimizer robustness while still relying on the quantum circuit to learn both components of the wavefunction. Our findings suggest this paradigm may be unnecessarily burdensome for near-term hardware. Demonstrated across a broad range of applications in this study, when the amplitude landscape is learned classically, and the quantum circuit is dedicated solely to learning the phase (sign) structure, optimization becomes measurably easier, gradients remain accessible at larger system sizes, and noise sensitivity decreases without sacrificing expressivity, as evidenced by sVQNHE's benign variational landscape and gradual transfer mechanism.

Importantly, the performance improvements observed across quantum simulation and combinatorial optimization arise from the division of labor itself. Amplitude learning thrives under classical updates, whereas phase learning benefits from quantum interference in shallow circuits. This explains why sVQNHE's scaling laws differ from VQE or QAOA: gradient variance no longer collapses exponentially, and measurement cost does not inflate with circuit depth or Hamiltonian complexity. In practice, this enables deeper phase-layer stacking on noisy hardware, with inherent error mitigation under realistic gate noise and finite sampling.

Empirically, in frustrated many-body problems like the 6-qubit J1-J2 model, sVQNHE reduces mean absolute energy error by $98.9\%$ and variance by $ 99.6\%$ relative to neural baselines, converging $\sim 19\times$ faster than hardware-efficient VQE. Similar gains hold for the 9-qubit 2D Heisenberg model and even sign-free systems like the 9-qubit TFIM or 12-qubit Ising, with error reductions up to $81.1\%$. Extending to molecular systems, sVQNHE yields superior ground-state energies and fidelities for a H2O molecule across geometries. For combinatorial optimization, it excels in MaxCut on Erdős–Rényi graphs, improving solution quality by up to $19\%$ and reducing measurements by $\sim 85\%$ in sampling-based tests up to 45 vertices.

More broadly, these findings shift how hybrid algorithms are conceptualized: the neural network as the primary learning module, while quantum hardware supplying essential but targeted phase transformations. This implies scaling quantum advantage via architectures where quantum roles are small yet critical, as seen in sVQNHE matching the classical Free-Energy Machine on a 1,485-vertex MaxCut instance and a 135-vertex MaxClique instance.
Overall, explicit quantum phase learning paired with classical amplitude modeling constitutes a principled hybrid strategy. By refining the quantum hardware's role, this work reinforces that quantum utility may depend more on complementary classical-quantum designs than on emulating universal circuits.

\section*{Methods}
\subsection*{The Hamiltonian of J1-J2 Model and Heisenberg Model}
For the 2D Heisenberg models are constructed with non-periodic boundary conditions, whose Hamiltonian can be written as, 
\begin{equation}\label{eq:HM}
H=h_i\sum_i Z^i+J_{ij} \sum_{edges}X^iX^j+Y^iY^j+Z^iZ^j.
\end{equation}
In our numerical simulation, we use the coefficient setting of $(h_i,J_{i,j})$ to be $(1,0.4)$ in $9$-qubit 2D Heisenberg model and $(1,1)$ in $3$-qubit Heisenberg model.

For the Hamiltonian of the n-qubit 1D J1-J2 spin model with non-periodic boundary conditions, it can be written as 
\begin{equation}
    H=\sum_{i=1}^{n-1} J_{1}(X^iX^{i+1}+Y^iY^{i+1}+Z^iZ^{i+1}) + \sum_{i=1}^{n-2} J_{2}(X^iX^{i+2}+Y^iY^{i+2}+Z^iZ^{i+2}).
\end{equation}
In our numerical simulation, we use the coefficients $J_1=1$ and $J_2=0.6$, since for the $J_2>0.5$ the sign problem is non-trivial for the J1-J2 model.

\subsection*{Detail of the Theoretical Framework}
\begin{theorem}\label{the:lie_ansatz}
    (Expressiveness of the Quantum Circuit Ansatz)\cite{larocca2023theory,allcock2024dynamical,holmes2022connecting,kazi2024analyzing,kiani2020learning}
    Let $ \mathfrak{g_1}$ and $\mathfrak{g_2}$ be the Lie algebras corresponding to the generators $\mathcal{G}_1$ and $ \mathcal{G}_2$, respectively, where:
\[
\mathcal{G}_1 = \{ Z_{D_{2j}} \mid j = 1, 2, \ldots, \lfloor \frac{m}{2} \rfloor\} \bigcup \{ Y_i \mid i = 1, 2, \ldots, n \},
\]
\[
\mathcal{G}_2 = \left\{ \sum Z_{D_{2j}} \mid j = 1, 2, \ldots, \lfloor \frac{m}{2} \rfloor \right\} \bigcup \left\{ \sum X_i \mid i = 1, 2, \ldots, n \right\},
\]
    where $m \in \{2,3,\ldots,n\}$, $X_i$ and $Y_i$ are the Pauli strings with Pauli X and Pauli Y operators at the ith position, respectively, $D_{2j}$ is the set of all 2j-combinations from the set $\{1, 2, \ldots, n\}$, and $Z_{D_{2j}}$ is the Pauli string with the Pauli Z operator on the set of positions $D_{2j}$. Then, the following statements hold:
    \begin{itemize}
        \item $ \mathfrak{g_1} = \mathfrak{su}(2^{n-1}) \oplus \mathfrak{su}(2^{n-1})$.
        \item $ dim(\mathfrak{g_2}) < dim(\mathfrak{g_1}) $.
    \end{itemize}
\end{theorem}

\begin{proposition}\label{propLayer} 
(Commuting Block Structure)
Let $M$ be a block diagonal matrix $M = M_{1,1} \oplus \cdots \oplus M_{m,m}$ with distinct eigenvalues across blocks. Then, any matrix commuting with $M$ must share its block structure.  
\end{proposition}

\begin{theorem}\label{the:landscape}
(Benign Landscape of sVQNHE) For an \(l\)-layer (\(l \geq 1\)) sVQNHE algorithm with a commuting diagonal ansatz \( W_l(\vec{\theta}) \), the cost function \(\mathcal{L}(\vec{\theta}) = \langle \psi(\vec{\theta}) | H | \psi(\vec{\theta}) \rangle\), where \(\psi(\vec{\theta})=Z_f F_l W_l(\vec{\theta})|\phi \rangle= W_l(\vec{\theta})|\phi_f \rangle\), exhibits at worst polynomially vanishing gradient variance \(\text{Var}[\partial_{\theta_k} \mathcal{L}] \in \Omega(1/\text{poly}(n))\) if the support set size of \(|\phi_f \rangle\) is polynomially bounded in \(n\). Here, \(Z_f\) is the normalization coefficient and \(|\phi_f \rangle=Z_f F|\phi \rangle\).
\end{theorem}
\noindent\textit{Proof}: 
To prove the theorem, we establish that the variance of the gradient \(\partial_{\theta_k} \mathcal{L}\) with respect to a parameter \(\theta_k\) in the $l$-th layer ansatz \( W_l(\vec{\theta}) \) scales at worst polynomially with the number of qubits \( n \), avoiding exponential vanishing\cite{cerezo2021cost}, due to polynomial constrained input and the commuting diagonal structure.

Since \( W_l(\vec{\theta}) \) is a diagonal unitary composed of commuting Z-related gates, it can be expressed as
\[
W_l(\vec{\theta}) = \sum_{x \in \{0,1\}^n} e^{i \xi(x, \vec{\theta})} |x\rangle\langle x|,
\]
where \(\xi(x, \vec{\theta}) = \sum_{k=1}^K \theta_k g_k(x)\) for basis functions \( g_k(x) \) (e.g. for the parameters $\theta_k$ of gate $R_{zz}$ applied on the $k_1,k_2$ qubit, $g_k(y)=(-1)^{y_{k_1} + y_{k_2}}$, where $y=|y_1 y_2 \ldots y_n\rangle$).

The gradient is:
\[
\partial_{\theta_k} \mathcal{L} = 2 \operatorname{Re} \left[ \langle \phi_f | (\partial_{\theta_k} W_l^\dagger(\vec{\theta})) H W_l(\vec{\theta}) | \phi_f \rangle \right],
\]
where
\[
\partial_{\theta_k} W_l(\vec{\theta}) = i \sum_x g_k(x) e^{i \xi(x, \vec{\theta})} |x\rangle\langle x|,
\]
so
\[
\partial_{\theta_k} W_l^\dagger(\vec{\theta}) = -i \sum_x g_k(x) e^{-i \xi(x, \vec{\theta})} |x\rangle\langle x|.
\]

To assess barren plateaus, compute the variance over random initializations \(\vec{\theta} \sim [0, 2\pi)^K\):
\[
\text{Var}[\partial_{\theta_k} \mathcal{L}] = \mathbb{E}_{\vec{\theta}}[(\partial_{\theta_k} \mathcal{L})^2] - (\mathbb{E}_{\vec{\theta}}[\partial_{\theta_k} \mathcal{L}])^2.
\]

The mean vanishes: \(\mathbb{E}_{\vec{\theta}}[\partial_{\theta_k} \mathcal{L}] = 0\), as \(\mathbb{E}_{\vec{\theta}}[e^{i \xi(x, \vec{\theta})}] = 0\) for non-constant \(\xi\) due to phase cancellation over \([0, 2\pi)\). Thus, the variance is:
\[
\text{Var}[\partial_{\theta_k} \mathcal{L}] = \mathbb{E}_{\vec{\theta}}[(\partial_{\theta_k} \mathcal{L})^2].
\]

Substitute the gradient
\[
\text{Var}[\partial_{\theta_k} \mathcal{L}] = \mathbb{E}_{\vec{\theta}}[\left[ \langle \phi_f | (\partial_{\theta_k} W_l^\dagger) H W_l | \phi_f \rangle \right]^2+\left[ \langle \phi_f | W_l^\dagger H (\partial_{\theta_k}W_l) | \phi_f \rangle \right]^2 + 2\langle \phi_f | (\partial_{\theta_k} W_l^\dagger) H W_l | \phi_f \rangle\langle \phi_f | W_l^\dagger H (\partial_{\theta_k}W_l) | \phi_f \rangle].
\]

Now, substitute the expressions:
\[
W_l(\vec{\theta}) = \sum_x e^{i \xi(x, \vec{\theta})} |x\rangle\langle x|,\quad \partial_{\theta_k} W_l(\vec{\theta}) = i \sum_y g_k(y) e^{i \xi(y, \vec{\theta})} |y\rangle\langle y|,\quad \partial_{\theta_k} W_l^\dagger(\vec{\theta}) = -i \sum_y g_k(y) e^{-i \xi(y, \vec{\theta})} |y\rangle\langle y|.
\]

Then:
\[
\langle \phi_f | (\partial_{\theta_k} W_l^\dagger) H W_l | \phi_f \rangle = -i \sum_{y,z} g_k(y) e^{-i \xi(y, \vec{\theta})} \langle \phi_f | y \rangle \langle y | H | z \rangle e^{i \xi(z, \vec{\theta})} \langle z | \phi_f \rangle,
\]
\[
\langle \phi_f | W_l^\dagger H (\partial_{\theta_k} W_l) | \phi_f \rangle = i \sum_{y,z} g_k(z) e^{-i \xi(y, \vec{\theta})} \langle \phi_f | y \rangle \langle y | H | z \rangle e^{i \xi(z, \vec{\theta})} \langle z | \phi_f \rangle.
\]
and
\begin{align}
&\langle \phi_f | (\partial_{\theta_k} V^\dagger) H V | \phi_f \rangle \cdot \langle \phi_f | V^\dagger H (\partial_{\theta_k} V) | \phi_f \rangle \nonumber\\
&= \sum_{y_1,z_1,y_2,z_2} g_k(y_1) g_k(z_2) e^{-i (\xi(y_1) - \xi(z_1))} e^{i (\xi(y_2) - \xi(z_2))} \langle \phi_f | y_1 \rangle \langle y_1 | H | z_1 \rangle \langle \phi_f | y_2 \rangle \langle y_2 | H | z_2 \rangle \langle z_1 | \phi_f \rangle \langle z_2 | \phi_f \rangle. \nonumber
\end{align}
Averaging over \(\vec{\theta}\):
\[
\mathbb{E}_{\vec{\theta}} \left[ e^{-i (\xi(y_1) - \xi(z_1) + \xi(y_2) - \xi(z_2))} \right] = \delta_{y_1 z_1} \delta_{y_2 z_2},
\]
because the expectation over \(\vec{\theta}\) cancels unless the phase differences are zero (e.g., \( y_1 = z_1 = y_2 = z_2 \)). 
The same argument applies to other items in $\text{Var}[\partial_{\theta_k} \mathcal{L}]$.

Thus we have
\[
\text{Var}[\partial_{\theta_k} \mathcal{L}] \geq \Omega( \sum_{y} g_k^2(y) |\langle y | H | y \rangle|^2 |\langle \phi_f | y \rangle|^4).
\]

For the form with restricted support set size, where \(|\langle \phi_f | y \rangle| > 0\) only at \(poly(n)\) positions—instead of the type where \(2^n\), by the Cauchy–Schwarz inequality we have:
\[
\text{Var}[\partial_{\theta_k} \mathcal{L}] \geq \Omega(\frac{1}{poly(n)}).
\]

Therefore, for $|\psi_f \rangle$ where the support set size is polynomially bounded, the barren plateau problem is not prone to occur.
$\hfill\square$

\vspace{5mm}
\noindent\textbf{Gradual transfer mechanism}\\
Our algorithm enforces strict classical-to-quantum information transfer by iteratively aligning the classically simulable $G_l$ (shallow quantum circuit blocks)  with the classical part $F_{l-1}$. This can be performed by different schemes, such as:\\
1. Frobenius norm minimization (global): Minimize the full matrix distance:  $|G_l - F_{l-1}|_F^2$.\\
2. Test state projection consistency (local): For a set of Haar-random test states $\{|\psi_{\text{test}}^k\rangle\}$, enforce consistency in the projected probability distribution
\begin{equation}
    P_G^k(s) = |\langle s|G_l|\psi_{\text{test}}^k\rangle|^2, \quad P_F^k(s) = |\langle s|F_{l-1}|\psi_{\text{test}}^k\rangle|^2
\end{equation}
by minimize expected KL divergence $\mathbb{E}_k\left[ \text{KL}(P_G^k \| P_F^k) \right] \to 0$. This test state sampling provides a weaker but practical consistency condition, avoiding direct optimization of $G_l $.

\subsection*{Measurement Strategy and Cost}
In our layered optimization strategy, we focus on performing measurements for any given layer.
To estimate the measurement cost of the Sign Ansatz, we use the parameter shift rule mentioned in Equation \eqref{eq:para shift} to compute the gradient of each parameter in the circuit. The parameter shift rule is defined as follows:
\begin{equation}\label{eq:para shift}
    \nabla_{\theta} f(x;\theta)=\frac{1}{2}[f(x;\theta+\frac{\pi}{2})-f(x;\theta-\frac{\pi}{2})].
\end{equation}
To maximize the efficiency of executing the parameter shift rule, we exploit the commutative property of $W(\theta^W)$. During iterative parameter updates, this approach simplifies the process: we only need to add the corresponding quantum gate after circuit W for each updated parameter. Specifically, for the parameters in $W(\theta^W)=W_1(\theta_1^W)\cdots W_{n_W}(\theta_{n_W}^W)$, we have 
\begin{equation}\label{eq:measure_1}
    \frac{\partial E_{P_i}(\theta^W)}{\partial \theta_k^W} = \frac{1}{2}(E_{P_i}(\theta_k^W+\frac{\pi}{2})-E_{P_i}(\theta_k^W-\frac{\pi}{2})).
\end{equation}
Since $W_k$ and $F$ are all commutative, and $E_{P_i}(\theta^W)=\bra{\psi_0}W^\dagger(\theta^W)F^\dagger P_i FW(\theta^W)\ket{\psi_0}$, we have
\begin{align}\label{eq:measure_2}
    &E_{P_i}(\theta_k^W \pm\frac{\pi}{2}) \nonumber\\
    &=\bra{\psi_0}W^\dagger(\theta^W)F^\dagger W_k^{\dagger}(\pm \frac{\pi}{2}) P_i  W_k(\pm \frac{\pi}{2})FW(\theta^W)\ket{\psi_0}.
\end{align}
Equation \eqref{eq:measure_1} and \eqref{eq:measure_2} imply that to obtain the gradient of $\theta^W$, we only need to add the measurement of Puali string $L_i=W_k^{\dagger}(\frac{\pi}{2}) P_i  W_k(\frac{\pi}{2})$. 

Note that any measurement of $Z \in L_i$ can be done by adding a factor $\prod_{i\in ind_z}(1-s_i)$ to the coefficient $(1-2s_0)$, where $ind_z$ is the position of Z in $L_i$ and $s_i$ is the Boolean value of the i index, so only changes in X and Y in $L_i$ will cause the number of measurements to increase. Let $P_i$ be a Pauli string containing exactly $|p_1|$ Pauli X or Y operators. Let $W_k$ be the rotation gate corresponding to $Z_p$ with $|p_2|$ operators. Since the transformations between X and Y can occur between $L_i$ and $P_i$ only if $|p_1 \cap p_2|$ is odd or $|p_1 \cap p_2| = |p_2|$, then for a Hamiltonian with $|p_1|\leq 2$, such as the 1D J1-J2 model, the total amount of measurement we need is $O(m_H)$, where $m_H$ is the number of Pauli strings $P_i$ in $H$. Here, $p_1$ and $p_2$ are non-empty subsets of the power set of all positions of the n qubit system. 

In summary, due to the commutation properties of the Sign Ansatz, our algorithm requires fewer measurements compared to general algorithms that rely on parameterized quantum circuits. For instance, considering the 1D J1-J2 model presented below \eqref{eq:J1-J2}, if $|p|\geq 2$ in the Sign Ansatz, the total number of measurements in a single iteration of our algorithm is equal to $3(n_1+n_2)+1$, where $n_1$ is the number of $\langle i,j \rangle$, and $n_2$ is the number of $\langle \langle i,j \rangle \rangle$. 
\begin{equation}\label{eq:J1-J2}
    H = J_1 \sum_{\langle i,j \rangle} (\sigma_i^x \sigma_j^x + \sigma_i^y \sigma_j^y + \Delta_1 \sigma_i^z \sigma_j^z)+ J_2 \sum_{\langle \langle i,j \rangle \rangle} (\sigma_i^x \sigma_j^x + \sigma_i^y \sigma_j^y + \Delta_2 \sigma_i^z \sigma_j^z) + B_H \sum_i \sigma_i^z.
\end{equation}
Here $\sigma^{x,y,z}$ are the Pauli matrices, $\Delta_1$ and $\Delta_2$ are the anisotropy parameters for nearest-neighbor and next-nearest-neighbor interactions, and $B_H$ is the strength of an external magnetic field.  $\langle i,j \rangle$ and $\langle \langle i,j \rangle \rangle$ denote sums over nearest-neighbor and next-nearest-neighbor spin pairs, respectively.

\subsection*{The details of our numerical simulation}
We use TensorCircuit\cite{zhang2023tensorcircuit} to efficiently obtain all the quantum circuits and the neural network numerical results with the following Ansatz setting, and the details of the VQNHE algorithm are in the \href{https://github.com/tencent-quantum-lab/tensorcircuit}{TensorCircuit Demo}.

$$\Qcircuit @C=0.5em @R=0.5em {
  & \gate{H} & \gate{Rz(\theta_1)} & \multigate{1}{Rzz(\theta_4)} & \qw & \qw & \qw & \sgate{Rzz(\theta_6)}{2} &\gate{Ry(\theta_7)} & \qw\\
  & \gate{H} & \gate{Rz(\theta_2)} & \qw & \multigate{1}{Rzz(\theta_5)} & \qw & \qw & \qw & \gate{Ry(\theta_8)} & \qw\\
  & \gate{H} & \gate{Rz(\theta_3)} & \qw & \qw & \qw & \qw & \gate{Rzz(\theta_6)} & \gate{Ry(\theta_9)} & \qw\\
  & & \vdots & & & & & & \vdots \\
& & &  &  &
}$$ 

The Sign Ansatz design equivalents for the sampling and noise versions of the numerical results with finite shots are detailed as follows. A combination of $YM$, $YP$, $C_z$, and $R_z$ gates equivalently replaces the $R_{zz}$ gate. Similarly, the $R_y$ gate is equivalently replaced by a combination of $XP$, $XM$, and $R_z$ gates. Here, $YM$ and $YP$ denote positive and negative rotations by $\frac{\pi}{4}$ around the Y-axis of the Bloch sphere, respectively, while $XM$ and $XP$ denote positive and negative rotations by $\frac{\pi}{4}$ around the X-axis of the Bloch sphere, respectively. The two-qubit gates used in our cases connect the edges of the model (1D J1-J2 model and 2D Heisenberg model). 

$$\Qcircuit @C=1.5em @R=0.7em {
&\gate{Ry(\theta_1)}&\gate{Rz(\theta_2)}& \ctrl{1} & \qw         & \qw         & \qw  \\
&\gate{Ry(\theta_3)}&\gate{Rz(\theta_4)}& \targ   & \ctrl{1}    & \qw         & \qw  & \times n\\
&\gate{Ry(\theta_5)}&\gate{Rz(\theta_6)}& \qw     & \targ       & \qw    & \qw \\
&\vdots & & & \vdots &\\
& & &  &  &
}$$ 

In the design of Hardware Efficient Ansatz, where the $CNOT$ is only connecting the two nearest qubits (2D Heisenberg Model), n is the number of repetitions. Note that after repeating the above block n times, one more layer of $R_y$ and one more layer of $R_z$ are added. For example, for HEA2 used in this paper, there are three $R_y$ layers, three $R_z$ layers, and two $CNOT$ layers. Similarly, the $CNOT$ gate in the numerical simulation of the finite-shot version is equivalently replaced by a combination of $YM$, $YP$, and $C_z$ gates.
\subsubsection*{Metrics}
In the experiment, we use the following metrics to evaluate the performance of the method:
\begin{itemize}
    \item Mean absolute error ratio $R_{MAE}$: 
    \begin{equation}\label{eq:R_MAE}
        R_{MAE}=\frac{MAE_1-MAE_0}{MAE_0}.
    \end{equation}
    The metric $R_{MAE}$ represents the relative reduction in mean absolute error (MAE) of different algorithms compared to the baseline algorithm (NN) with the same number of iterations. Here $MAE_0$ represents the MAE of the baseline algorithm (NN) being compared. The smaller the $R_{MAE}$ value, the better the compared algorithm is relative to NN.
    \item Variance ratio $R_{Var}$:
    \begin{equation}\label{eq:R_var}
        R_{Var}=\frac{Var_1-Var_0}{Var_0}.
    \end{equation}
    The metric $R_{Var}$ represents the relative reduction in the variance of different algorithms compared to the baseline algorithm (NN) given the same number of iterations. Among them, $Var_0$ represents the variance of the baseline algorithm (NN) being compared. The smaller the $R_{Var}$ value, the more stable the compared algorithm is relative to NN, and the smaller the variability of the results.
    \item Estimated approximation ratio $R_e$ for the MaxCut problem:
    \begin{equation}\label{eq:R_e}
        R_e=\frac{MaxCut_{algo}}{MaxCut_{max}}.
    \end{equation}
    The metric $R_e$ provides an estimated approximation ratio based on the best-known result for the MaxCut problem, which measures the relative solution quality. Here, $MaxCut_{max}$ represents the maximum MaxCut value achieved across all evaluated algorithms, while $MaxCut_{algo}$ denotes the MaxCut value obtained by the specific algorithm under assessment. A value of $R_e$ approaching $100\%$ indicates superior performance relative to the other algorithms compared.
\end{itemize}
\subsubsection*{Parameters used in MLP under different models}

To ensure fair comparison and reproducibility, we specify the architecture of the MLP used in each model. The MLP serves as the amplitude-generating component in sVQNHE and VQNHE, and as the standalone ansatz in the baseline NN method.  The width is defined relative to the number of qubits $n$, and depth refers to the number of hidden layers (input and output layers excluded).

\begin{table}[htp!]
\centering
\small
\scalebox{0.9}{
\begin{tabular}{@{}lcccc@{}}
\toprule
\textbf{Model} & \textbf{MLP Type} & \textbf{Depth} & \textbf{Width} & \textbf{Notes} \\
\midrule
1D J1-J2 & Real-valued & 5 & $3n$ &  \\
2D Heisenberg & Real-valued & 2 & $3n$ & \\
1D TFIM & Real-valued & 2 & $3n$ & \\
Ising & Real-valued & 2 & $3n$ & \\
1D Heisenberg & Non-negative real & 2 & $3n$ & Absolute value output for amplitude \\
Water molecule ($H_2O$, 10 qubits) & Real-valued & 3 & $6n$ &  \\
MAXCUT ($\leq 45$ vertices) & Non-negative real & 2 & $3n$ & $n$: \# qubits (after encoding) \\
MAXCUT (1485 vertices) & Real-valued & 2 & $10n$ & Increased width for large-scale graphs \\
\bottomrule
\end{tabular}}
\caption{\small Summary of MLP architectures used across different physical and combinatorial problems.}
\label{tab:MLP}
\end{table}


\section*{Acknowledgements}
CYH acknowledges support from National Ke Research and Development Program of China (2024FA1306400), Natural Science Foundation of China (22373085). AH gratefully acknowledges the sponsorship from the National Natural Science Foundation of China (NSFC) (Grant No. 62541160274), City University of Hong Kong (Project No. 7006103), CityU Seed Fund in Microelectronics (Project No. 9229135), and Hon Hai Research Institute (Project No. 9231594, 9239182). YYS acknowledges support from the National Natural Science Foundation of China (Grant No. 12404575), the Shandong Provincial Natural Science Foundation (Grant No. ZR2022LLZ008), the Natural Science Foundation of Shanghai (Grant No. 23ZR1469600), and the Shanghai Sailing 
Program (Grant No. 23YF1452600).


\section*{Code and data availability}
The source code and data associated with this project are publicly available. They can be accessed at the following GitHub repository: https://github.com/renmengzhen/sVQNHE

\section*{Supplementary Information}
\subsection*{Optimizing Quantum Resources with Sign Ansatz VQNHE}
\begin{figure}[htp!]
\centering
\includegraphics[width=0.8\textwidth]{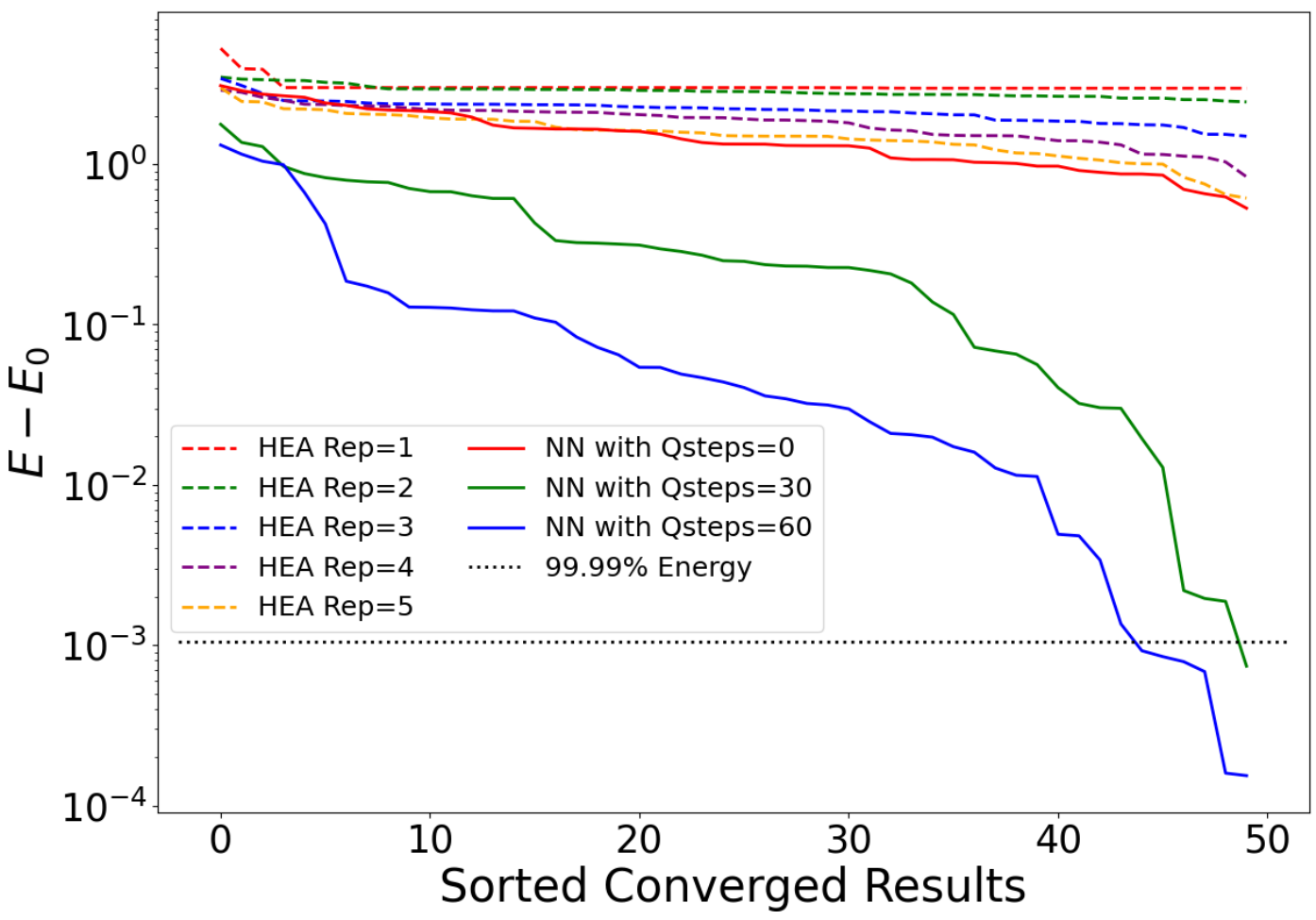}
\caption{9-qubit Heisenberg Model Result. The x-axis shows sorted results by the converged energy with different random initial parameters. The y-axis is the energy difference from the ground state. The red, green, and blue dashed lines are responding to Hardware Efficient Ansatz with repetitions equal to 1, 2, and 3, and the solid line is Complex MLP(3,3) with Sign Ansatz, where the total number of parameters in the neural network is 317. Both of them start from the initial state $\ket{+...+}$. The dotted line $99.99\%$ Energy is 0.0001 times the ground state energy.} \label{fig3}
\end{figure}

In this section, we evaluate the quantum resource efficiency of the sVQNHE framework against the Hardware Efficient Ansatz VQE. First, we test different ansatz on the 9-qubit 2D Heisenberg model with the initial state $\ket{++...++}$. The results are shown in Figure \ref{fig3}. It can be seen that sVQNHE uses very few quantum resources to converge to the desired accuracy compared to both neural networks and VQE, which implies that the sVQNHE reaches the desired accuracy with fewer quantum resources compared to VQE in the number of circuits, and the number of parameters. Besides, since the zero quantum resource results indicate that without the assistance of Sign Ansatz, the neural network itself cannot reach good accuracy, we can secondly claim that this framework can greatly improve the expressiveness of the neural network with limited quantum resources. In addition, we also verified the quantum resource efficiency of sVQNHE relative to VQE on the J1-J2 model. The results are shown in the Table \ref{tab:convergence_efficiency}.

\begin{table}[htp!]
\centering
\caption{Convergence Efficiency Comparison for the 6 qubit J1-J2 Model}
\label{tab:convergence_efficiency}
\begin{tabular}{@{}cccc@{}}
\toprule
\textbf{Method} & \textbf{Average Optimization Steps} & \textbf{Success Probability (\%)} & \textbf{Summary} \\
\midrule
sVQNHE & 132.67 & 60 & faster convergence with higher reliability \\
VQE (2-layer HEA) & 1676.00 & 5 & more steps and has a lower success rate \\
\bottomrule
\end{tabular}
\caption*{Explanation: This table compares the convergence efficiency of sVQNHE and VQE (with a 2-layer hardware-efficient ansatz) for the 6-qubit J1-J2 model. The metrics include the average number of optimization steps required to reach 99.45\% of the ground state energy and the probability of success. sVQNHE outperforms VQE by requiring fewer steps (132.67 vs. 1676) and achieving a higher success rate (60\% vs. 5\%), demonstrating superior efficiency in quantum resource utilization.}
\end{table}

\subsection*{Sampling Efficiency Verification}
This section demonstrates the rationale for sampling bitstrings directly from the quantum circuit component—specifically, from the L-layer phase/symbol hypothesis—rather than from a classical neural network that approximates the amplitude distribution. Beyond the inherent potential quantum advantage associated with sampling from the symbol hypothesis itself, this choice also offers clear numerical benefits during hybrid quantum-classical optimization.

\begin{figure}[htp!]
\centering
\includegraphics[width=0.9\textwidth]{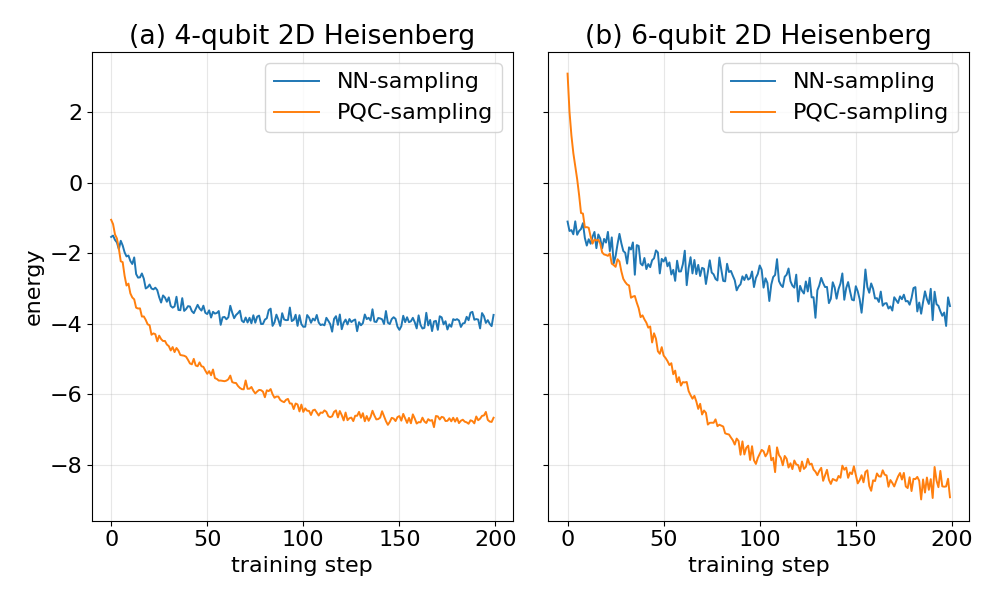}
\caption{Sampling Efficiency Verification under 2D Heisenberg Hamiltonian. (a) Convergence curves for the 4-qubit case using NN-Sampling (bitstrings generated by the classical neural network) versus PQC-Sampling (direct sampling from the quantum circuit component). (b) Corresponding results for the 6-qubit case.  The horizontal axis represents training iterations, and the vertical axis represents the obtained variational energy (lower is better). All simulations use an identical number of shots.} \label{fig:samp_nn}
\end{figure}

As shown in Fig. \ref{fig:samp_nn}, we compare the convergence behavior of the sVQNHE model under two distinct sampling strategies. In the NN-Sampling approach, bitstrings are generated by the classical neural network component and subsequently evaluated by the quantum circuit. In contrast, the standard sVQNHE scheme (referred to as PQC-Sampling) samples bitstrings directly from the quantum component, followed by neural-network-based weighting of the samples. To eliminate any possible advantage arising from difficult-to-simulate distributions in multi-layer circuits, we restrict the comparison to a single-layer (first-layer) sVQNHE ansatz. For each sampling strategy, we perform three independent training runs of 200 steps each under the same number of shots and select the evolution path that achieves the lowest final energy for comparison.

The results clearly indicate that direct sampling from the quantum circuit substantially outperforms NN-Sampling. Under identical computational budgets (shots=1200), quantum-component sampling yields markedly more stable training trajectories, faster overall convergence, and an increasingly pronounced performance advantage as system size grows. These findings suggest that sVQNHE can efficiently generate high-quality samples, thereby enabling exploration of a more representative and effective solution space.

\subsection*{Preliminary Resource Consumption Analysis for Larger Instances}
\begin{table}[htp!]
\centering
\caption{Resource Comparison for Larger Instances}
\label{tab:resource_comparison}
\begin{tabular}{@{}cccccc@{}}
\toprule
Qubits & Encoding & Vertices & \begin{tabular}[c]{@{}c@{}}brickwork-VQE\\ measurement cost\end{tabular} & Comparison & \begin{tabular}[c]{@{}c@{}}sVQNHE\\ measurement cost\end{tabular} \\ \midrule
\multirow{2}{*}{17} & 2 & 408 & 3000 & $>$ & 409 \\
 & 3 & 2040 & 24150 & $>$ & 4081 \\
\multirow{2}{*}{30} & 2 & 1305 & 9612 & $>$ & 1306 \\
 & 3 & 12180 & 141246 & $>$ & 24361 \\ \midrule
\end{tabular}
\caption*{Explanation: This table compares the measurement costs of sVQNHE and brickwork-VQE under 2nd and 3rd order polynomial encodings for 17 and 30 qubits. The vertex count represents the maximum number of graph vertices for the maxcut problem corresponding to each qubit count and encoding scheme. Measurement cost reflects the number of measurement circuits required per iteration. sVQNHE consistently requires fewer circuits, demonstrating superior efficiency in quantum resource utilization.}
\end{table}

In this section, we compare the measurement requirements for a single iteration of the brickwork-VQE and sVQNHE schemes under 2nd and 3rd order polynomial encodings \cite{sciorilli2025towards}, with equal shots per operator. As shown in the Table. \ref{tab:resource_comparison}, the savings in measurements, defined as the reduction when using sVQNHE compared to brickwork-VQE, scale linearly with the number of vertices in the graph.  While this scaling is linear, the absolute savings become significantly large for graphs with many vertices, resulting in a substantial efficiency advantage for sVQNHE in large-scale quantum systems.

\bibliography{reference}

\end{document}